\newcommand{\s}{\sigma}
\newcommand{\g}{\Gamma}
\newcommand{\m}{\mu}

\documentclass[oneside,a4paper,12pt]{amsart}
\usepackage{amsmath}               
\usepackage{amscd}
\usepackage{amsthm}                
\usepackage{times}
\usepackage{graphicx}
\usepackage{psfrag}

\numberwithin{equation}{section}

\title[Spherical shells]%
{The general solution for relativistic spherical shells }

\author[J. Kijowski]{J. Kijowski}
\address{Institute for Theoretical Physics, \hfill\break\indent Polish Academy of Sciences, Warsaw, Poland
and \hfill\break\indent Department of Mathematical Methods in
Physics, \hfill\break\indent University of Warsaw, Poland}
\email{kijowski@theta1.cft.edu.pl}
\author[G. Magli]{G. Magli}
\address{Dipartimento di Matematica,\hfill\break\indent Politecnico di
Milano, Italy} \email{magli@mate.polimi.it}
\author[D. Malafarina]{D. Malafarina}
\address{Dipartimento di Matematica,\hfill\break\indent Politecnico di
Milano, Italy and \hfill\break\indent Institute for Theoretical
Physics, \hfill\break\indent Polish Academy of Sciences, Warsaw,
Poland} \email{malafarina@mate.polimi.it}

\begin{document}
\swapnumbers
\theoremstyle{plain}\newtheorem{teo}{Theorem}[section]
\theoremstyle{plain}\newtheorem{prop}[teo]{Proposition}
\theoremstyle{plain}\newtheorem{lem}[teo]{Lemma}
\theoremstyle{plain}\newtheorem{cor}[teo]{Corollary}
\theoremstyle{definition}\newtheorem{defin}[teo]{Definition}
\theoremstyle{remark}\newtheorem{rem}[teo]{Remark}
\theoremstyle{plain} \newtheorem{assum}[teo]{Assumption}
\theoremstyle{definition}\newtheorem{example}[teo]{Example}

\begin{abstract}

The general exact solution of the Einstein-matter field equations
describing spherically symmetric shells satisfying an equation of
state in closed form is discussed under general assumptions of
physical reasonableness. The solutions split into two classes: a
class of ``astro-physically interesting" solutions describing
``ordinary" matter with positive density and pressure, and a class
of ``phantom-like" solutions with positive density but negative
active gravitational mass, which can also be of interest in
several ``very strong fields" regimes. Known results on
linear-barotropic equations of state are recovered as particular
cases.

\end{abstract}

\maketitle

\section{Introduction}

Relativistic shells have found applications in an impressive list
of different scenarios, such as sources of vacuum gravitational
fields \cite{Berezin}-\cite{DeLaCruz2}, cosmological models
\cite{Sato1} and \cite{Lake}, gravitational collapse
\cite{Lindblom}, canonical quantization of gravity
  \cite{JK1} and \cite{JK2}, higher dimensional models in brane world, quantum gravitational collapse \cite{H},
entropy of black holes, motion of thin clouds surrounding an
exploding star \cite{Nun1}, sources of ``phantom" fields and dark
energy \cite{phantom}. In spite of the wide range of their
applications, very few analytical solutions in closed form are
known even in spherical symmetry. Exceptions are the dust (i.e.
zero pressure) shells, which were completely analyzed already in
the pioneering work by Israel \cite{Israel1}, \cite{Israel2}, and
solutions with linear-barotropic equations of state of the form
$P=\Gamma_0 \s$ with a constant $\Gamma_0$. In the present paper
we discuss the general exact solution for shells satisfying an
equation of state in closed form $P=\Gamma(\s)$ with some general
assumption of physical reasonableness on the behavior of the
function $\Gamma$. Our solutions include as particular cases the
linear e.o.s. cited above but also, for instance, non-linear
polytropic equations of state of the form
$\Gamma(\s)=\Gamma_0\s^\nu$ \cite{KMM}. It turns out that the
qualitative behavior of spherical shells always presents two
``branches", an ``astro-physically relevant" branch in which the
weak energy condition is strictly satisfied, and a ``phantom"
branch in which the density remains positive but the active
gravitational mass $\s+P$ is strictly negative.

\section{The field equations }

We consider a spherical shell in vacuum. The shell ``separates" a
portion of Minkowski space (``interior") from Schwarzschild space
of arbitrarily fixed mass parameter $M$ (``exterior"). The field
equations relate the jump of the exterior curvature with the
matter content of the shell, i.e. a surface distribution of
stress-energy. It can be shown \cite{Nun2} that such equations can
be written as follows:
\begin{equation}\label{conser}
  \dot{\sigma}=-2\frac{\dot{R}}{R}\left(\sigma+P\right)
\end{equation}
\begin{equation}\label{Rpunto}
  \dot{R}^2=V(R,\s)
\end{equation}
where $R(t)$ is the ``radius" of the shell as a function of the
proper time on the shell, $\s$ and  $P$ are the surface
(rest-frame) energy-density and the surface pressure (due to
spherical symmetry, the two eigenvalues of the stress are equal,
i.e. the material is necessarily a ``two-dimensional perfect
fluid") and the ``effective potential" $V$ is given by
$$
V(R,\s):=\frac{M^2}{16\pi^2 R^4\sigma^2}+4\pi^2 R^2\sigma^2-1
  +\frac{M}{R}.
$$

The equations (\ref{conser}) and (\ref{Rpunto}) contain three
unknowns, and become a closed set once a further relation has been
given. This has been achieved in various ways in the literature;
for instance, one can assign the pressure profile as a ``a priori"
known function \cite{Nun1} and calculate the corresponding density
profile. However, in this way, the resulting ``equation of state",
i.e. the relationship between density and pressure which holds
during the dynamics may turn out to be unphysical. Here, we are
going to consider the case of a closed equation of state of the
barotropic type, namely $P=\g (\s)$. In this case, as we shall
see, the equations above decouple and the dynamics can be
qualitatively analyzed.

The function $\g$ has to satisfy a certain set of assumptions
which assure the physical reasonability of the model. It seems
that a reasonable compromise between generality and physical
reasonableness can be obtained assuming that $\g (\s )$ is a
continuous, monotonic function, differentiable except perhaps at
$\s =0$, where its value vanishes. These assumptions comprise a
wide range of equations of state and, in particular, the
linear-barotropic ($P=\g_0\s$ with a constant $\g_0$) that has
been widely studied in the literature \cite{Goncalves},
\cite{Zlo}, and the polytropic case $P=\g_0\s^\nu$ (both these
e.o.s. reduce to the newtonian polytropes in the weak field regime
\cite{Tooper1}, \cite{Tooper2}).

If the weak energy condition is required, then solutions have to
be selected in such a way that $\s>0$, and on such solutions the
effective gravitational mass
$$
\m(\s):=\s+\g (\s )
$$
must be non negative. Therefore the requirement $\g(\s)\geq -\s$
has to be added. However, shells with positive density but
negative effective mass are of interest on their own, for instance
as toy models for sources of dark energy (``phantom'' fields, see
e.g. \cite{phantom}, exotic matter and wormholes, see
\cite{lobo}), thus we shall consider also
this possibility in what follows.\\

\section{The structure of the solutions }

We take initial data $\s(R_0)=\s_0>0$ and consider the behavior of
the solutions of the ordinary differential equation
\begin{equation}\label{con}
  \dot{\sigma}=-2\frac{\dot{R}}{R}\m(\s).
\end{equation}

First of all, we notice that it can be reduced to a quadrature
\begin{equation}\label{cons}
 \int_{\sigma_0}^{\sigma(R)}\frac{d\sigma}{\m(\s)}=-2\log (R/R_0).
\end{equation}

Thus, if the function $1/\m(\s)$ is integrable in neighborhood of
$+\infty$, the density diverges then this occurs at a finite non
zero value $R_s$ (with $R_0>R_s>0$) (``Case A"). If instead the
function $1/\m(\s)$ is not integrable in neighborhood of
$+\infty$, the density can diverge only as $R$ approaches zero
(``Case B"). Further, a straightforward analysis shows that the
following cases have to be distinguished:\\

Case I): $\m(\s)>0$ for any $\s$:

In this case each solution is strictly decreasing from $+\infty$
to zero (one among behaviors A or B occurs). Weak energy condition
is always satisfied. Since $\s=0$ is a solution, if $d\g/d\s$ is
continuous in $\s=0$ the standard uniqueness theorem for ODE gives
$\s \to 0$ as $R\to +\infty$. For completeness we notice, however,
that one could also conceive models in which $d\g/d\s$ has a
finite jump in $\s=0$. In this case a loss of uniqueness occurs at
a finite value $R_l$ (at which the solution goes to zero and can
be prolonged to be zero, thus giving a sort of ``evaporation" of
the shell). We thus have four qualitatively different cases
reported in figures 1-4.\\

Case II): There exist a zero (at $\s_p$) of the function $\m(\s)$:

Clearly $\s=\s_p$ is a regular solution at which no loss of
uniqueness can occur. Thus the behavior depends on the initial
datum for $\s$:

(IIa) If $\s_0>\s_p$ then all solutions are strictly decreasing
and approach from above $\s_p$ as $R\to +\infty$; one among cases
A and B occurs. Weak energy condition is satisfied everywhere.

(IIb) If $\s_0<\s_p$ then all solutions are strictly increasing
and approach $\s_p$ from below as $R\to +\infty$; their behavior
``in the past" intersects $\s=0$ where a loss of uniqueness
occurs. Weak energy condition is violated everywhere because the
density is positive but $\m$ is strictly negative. Two behaviors
must still be distinguished: if the derivative
$\s''=-\s'/R(3+d\g/d\s)$ has no zeroes, the solutions come out
from zero with infinite tangent, if it has a zero the solutions
have  a flex point, and the matching at $\s=0$ is smooth in this
case. All in all we have the four qualitatively different cases
reported in figures 5-8.

\section{Discussion and conclusion}

Once the behavior of the energy function $\sigma$ is known the
corresponding qualitative behavior of the shell radius $R(t)$ can
be read off from the ``effective potential" equation
(\ref{Rpunto}).

First of all, we notice that, due to the presence of the
$M/R+4\pi^2 R^2\sigma^2$ term, the potential always diverges at
the left end side of the interval of definition, in all the eight
qualitatively different cases discussed above. The behavior at the
other extreme is divergent as well in the cases I without loss of
uniqueness and in all cases II (where $V$ diverges as $R$ goes to
infinity). Thus in all such cases the potential is a continuous
curve positively diverging at the extremes, and it follows that it
has an absolute minimum. The dynamics is allowed in both the
neighborhoods of the extremes, and it is allowed everywhere if $V$
is positive at the minimum. Bouncing or also oscillating behaviors
can appear as well, if the minimum is negative and therefore there
is at least one region in which the dynamics is not allowed.

Shells with general equation of state can thus be viewed as very
useful ``toy" models for astrophysical objects, in order to
investigate the effect of the equation of state on the dynamics in
several physically interesting scenarios. Work in this direction
is in progress.

 \begin{center}
\begin{figure}[hhh]
  \psfrag{x}{$R$}
  \psfrag{y}{$\sigma$}
  \psfrag{s}{(IA)}
  \psfrag{R}{$R_s$}
  \psfrag{1}{}
  \psfrag{2}{}
  \psfrag{3}{}
  \psfrag{4}{}
  \psfrag{5}{}
  \psfrag{6}{}
  \psfrag{7}{}
  \psfrag{8}{}
  \psfrag{10}{}
  \includegraphics{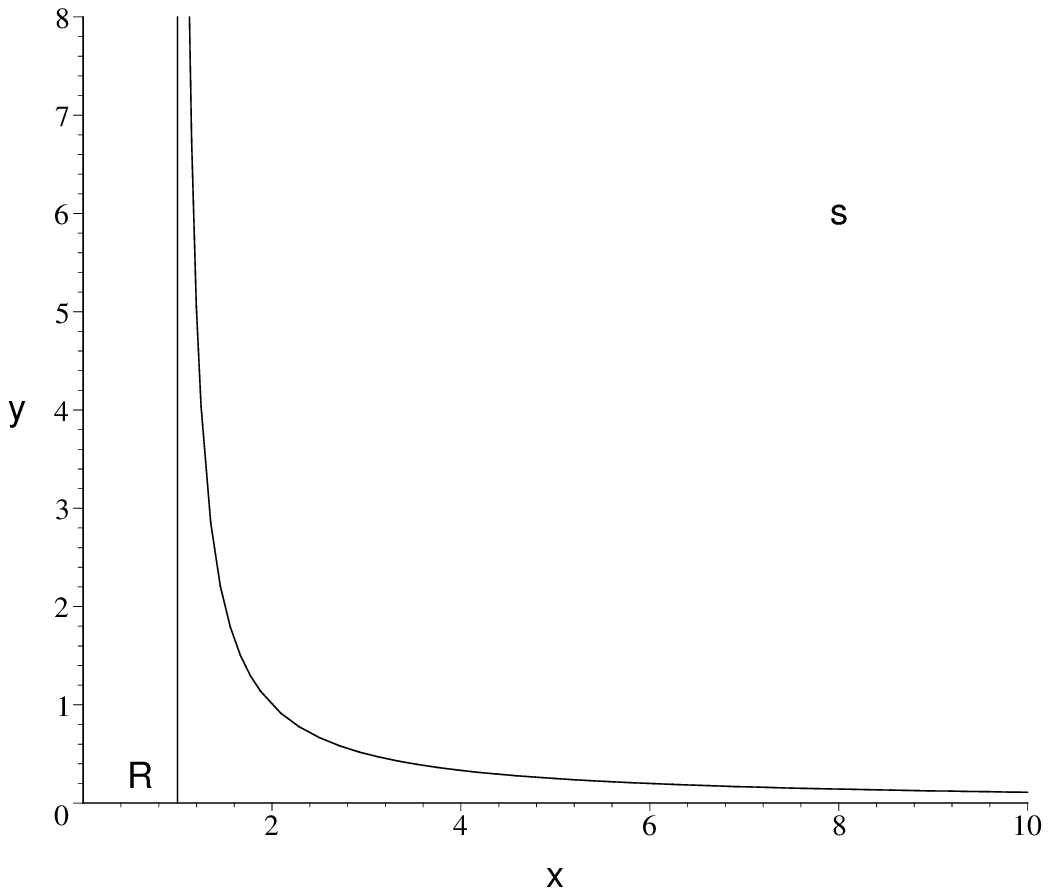}
\caption{$\sigma(R)$: Case IA when $\frac{d\Gamma}{d\sigma}$ is
continuous in $\sigma=0$.}\label{fig:IaA}
  \psfrag{s}{(IB)}
  \includegraphics{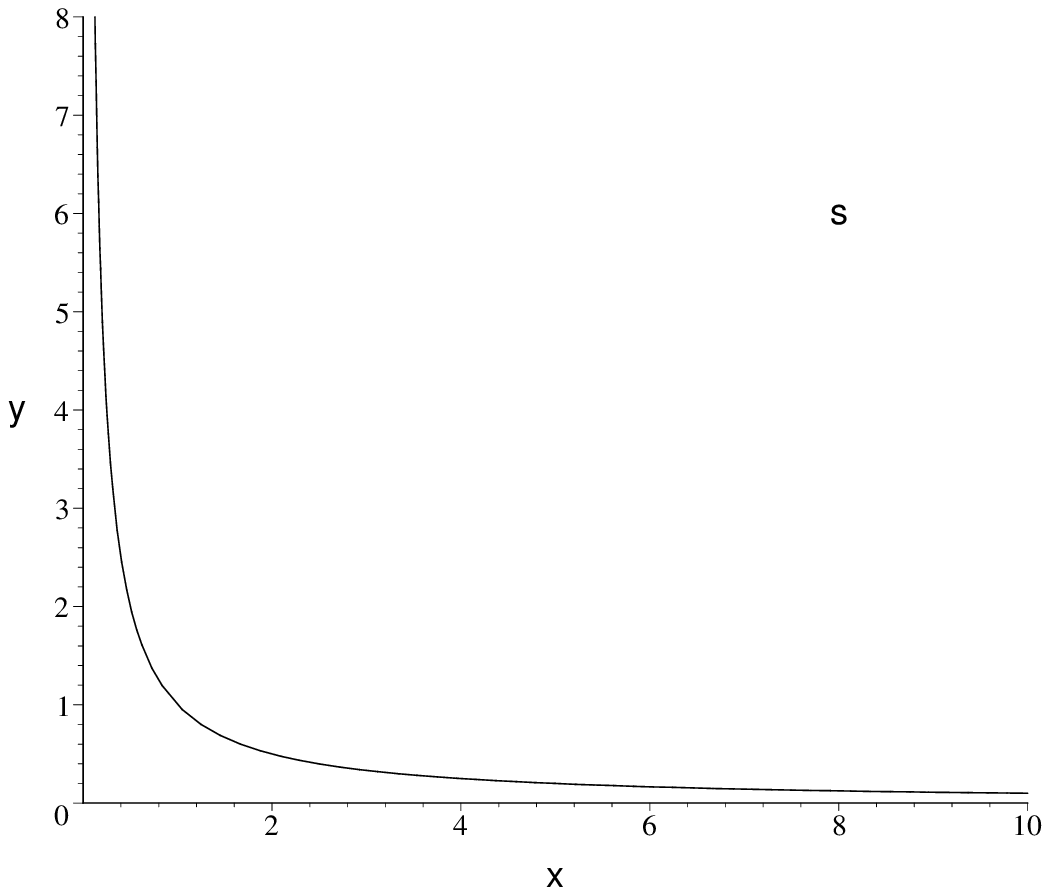}
  \caption{$\sigma(R)$: Case IB when $\frac{d\Gamma}{d\sigma}$ is
continuous in $\sigma=0$.}\label{fig:IaB}
\end{figure}
\end{center}

\begin{center}
\begin{figure}[hhh]
  \psfrag{x}{$R$}
  \psfrag{y}{$\sigma$}
  \psfrag{s}{(IA)}
  \psfrag{R}{$R_s$}
  \psfrag{Q}{$R_l$}
  \psfrag{0.5}{}
  \psfrag{1}{}
  \psfrag{1.5}{}
  \psfrag{2}{}
  \psfrag{2.5}{}
  \psfrag{3}{}
  \psfrag{4}{}
  \psfrag{6}{}
  \psfrag{8}{}
  \psfrag{10}{}
  \includegraphics{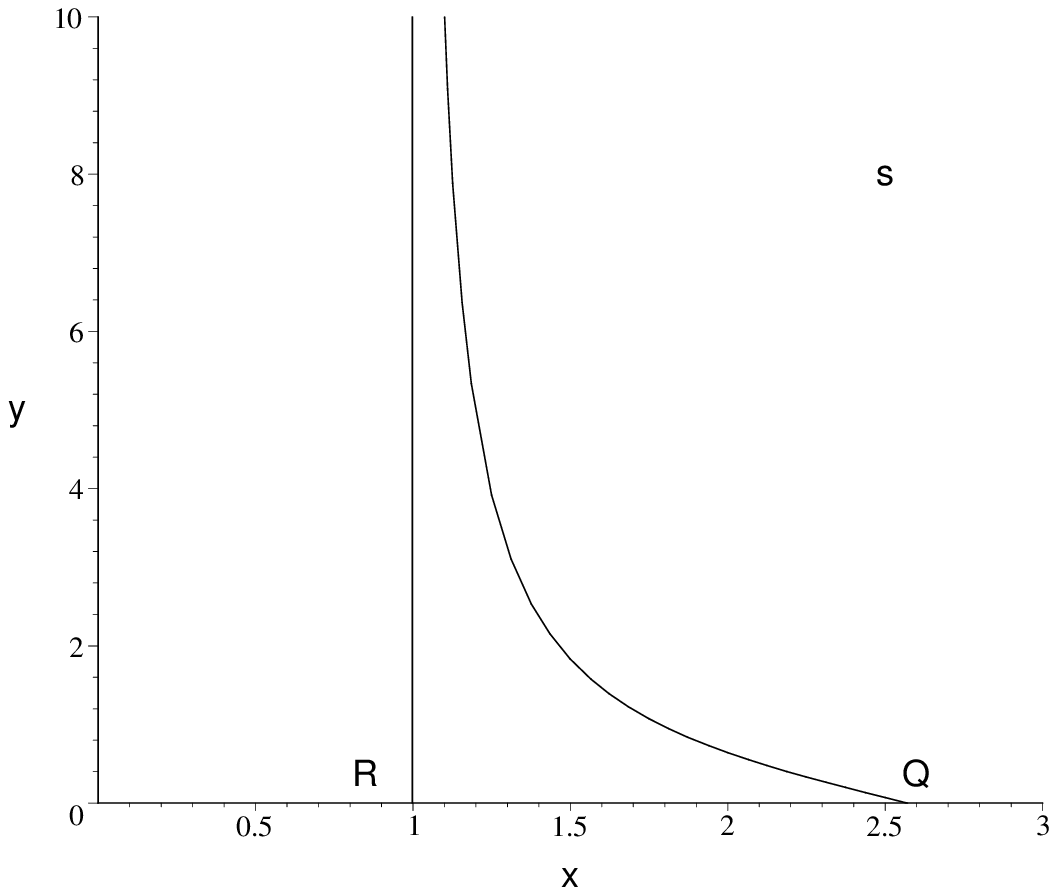}
\caption{$\sigma(R)$: Case IA when $\frac{d\Gamma}{d\sigma}$ is
discontinuous in $\sigma=0$.}\label{fig:IbA}
  \psfrag{s}{(IB)}
  \psfrag{R}{$R_l$}
  \psfrag{0.2}{}
  \psfrag{0.4}{}
  \psfrag{0.6}{}
  \psfrag{0.8}{}
  \psfrag{1.2}{}
  \psfrag{1.4}{}
  \psfrag{1.6}{}
  \psfrag{1.8}{}
  \includegraphics{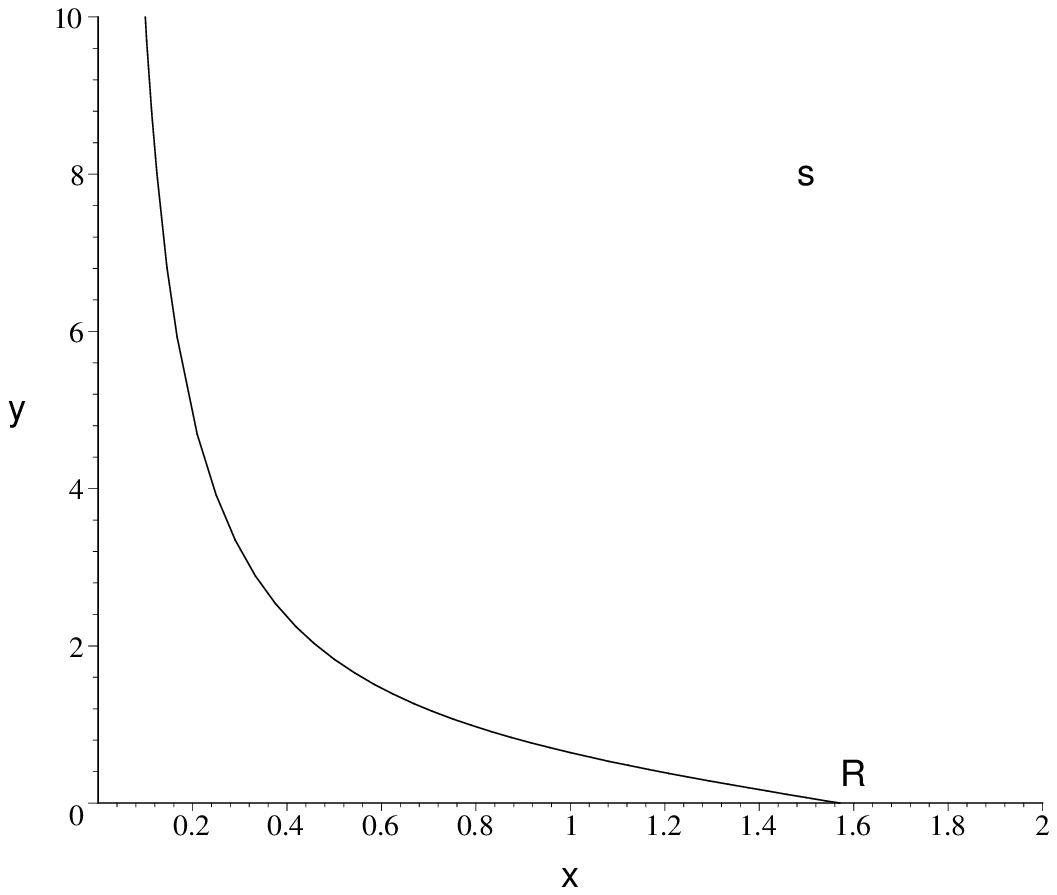}
  \caption{$\sigma(R)$: Case IB when $\frac{d\Gamma}{d\sigma}$ is
discontinuous in $\sigma=0$.}\label{fig:IbB}
\end{figure}
\end{center}

\begin{center}
\begin{figure}[hhh]
  \psfrag{x}{$R$}
  \psfrag{y}{$\sigma$}
  \psfrag{s}{(IIaA)}
  \psfrag{R}{$R_s$}
  \psfrag{sp}{$\sigma_p$}
  \psfrag{1}{}
  \psfrag{2}{}
  \psfrag{3}{}
  \psfrag{4}{}
  \psfrag{5}{}
  \psfrag{6}{}
  \psfrag{7}{}
  \psfrag{8}{}
  \psfrag{10}{}
  \includegraphics{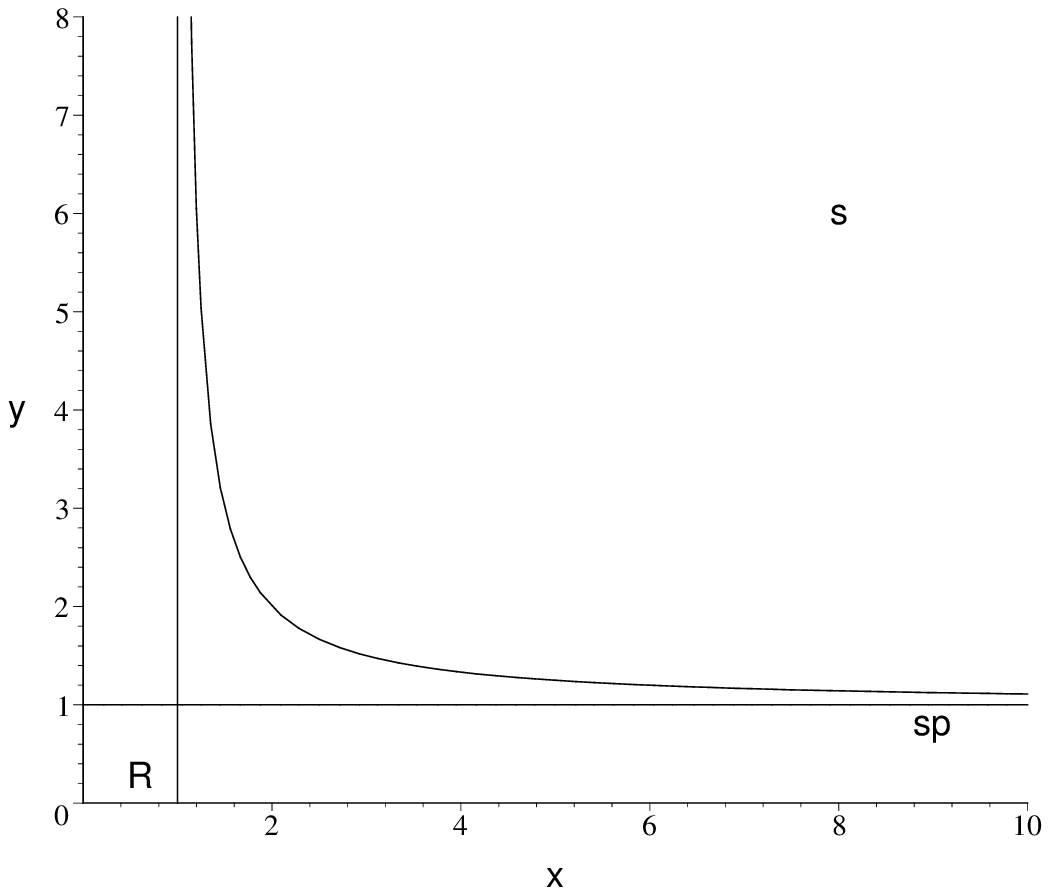}
\caption{$\sigma(R)$: Case IIaA.}\label{fig:IIaA}
  \psfrag{s}{(IIaB)}
  \includegraphics{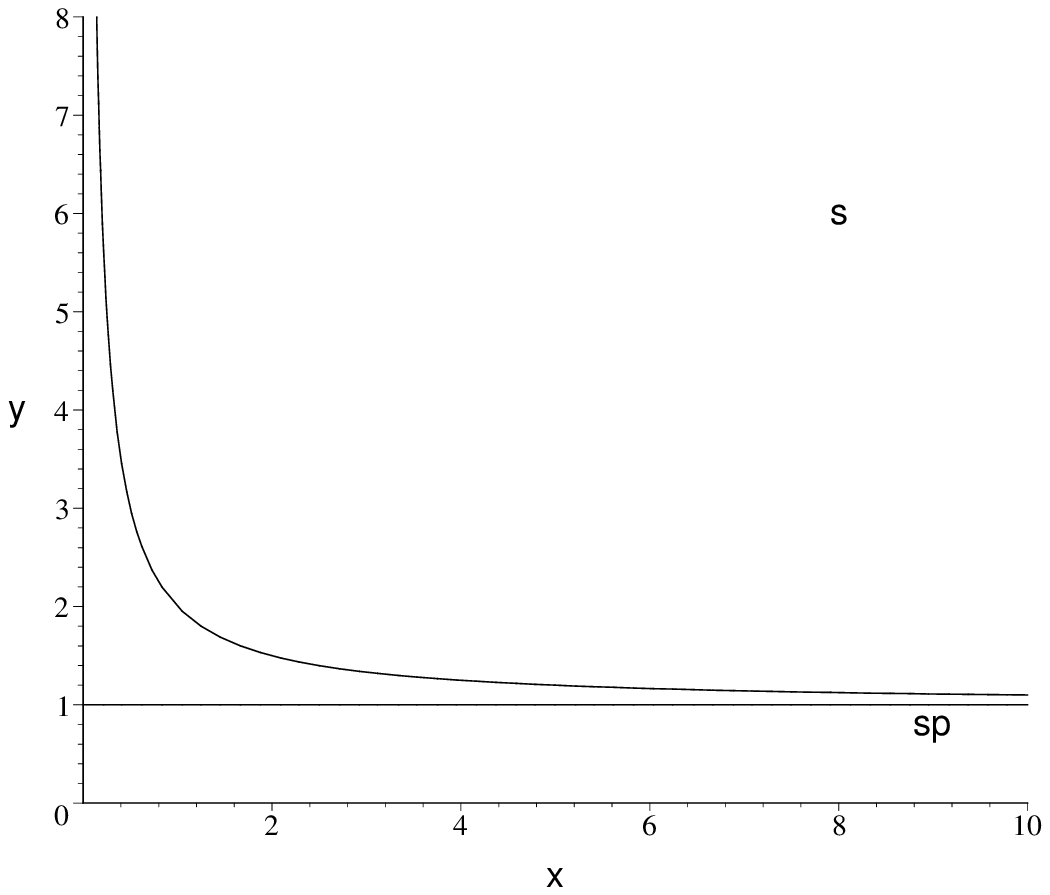}
  \caption{$\sigma(R)$: Case IIaB.}\label{fig:IIaB}
\end{figure}
\end{center}

\begin{center}
\begin{figure}[hhh]
  \psfrag{x}{$R$}
  \psfrag{y}{$\sigma$}
  \psfrag{s}{(IIb$\alpha$)}
  \psfrag{R}{$R_s$}
  \psfrag{sp}{$\sigma_p$}
  \psfrag{0.2}{}
  \psfrag{0.4}{}
  \psfrag{0.6}{}
  \psfrag{0.8}{}
  \psfrag{1.2}{}
  \psfrag{1.4}{}
  \psfrag{1.6}{}
  \psfrag{1.8}{}
  \psfrag{1}{}
  \psfrag{2}{}
  \psfrag{3}{}
  \psfrag{4}{}
  \psfrag{5}{}
  \psfrag{6}{}
  \psfrag{7}{}
  \psfrag{8}{}
  \psfrag{10}{}
  \includegraphics{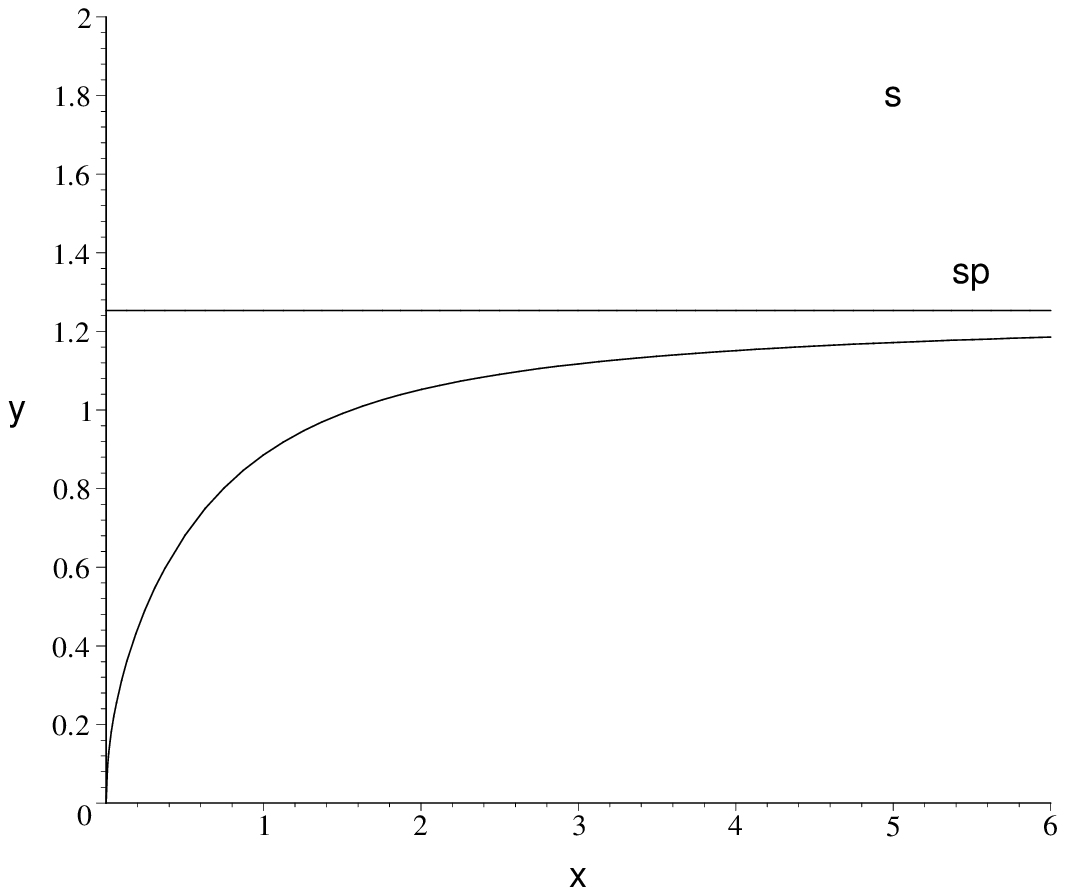}
\caption{$\sigma(R)$: Case IIb when $\sigma''$ has no
zeroes.}\label{fig:IIbAlfa}
  \psfrag{s}{(IIb$\beta$)}
  \psfrag{12}{}
  \psfrag{14}{}
  \psfrag{16}{}
  \psfrag{18}{}
  \psfrag{20}{}
  \includegraphics{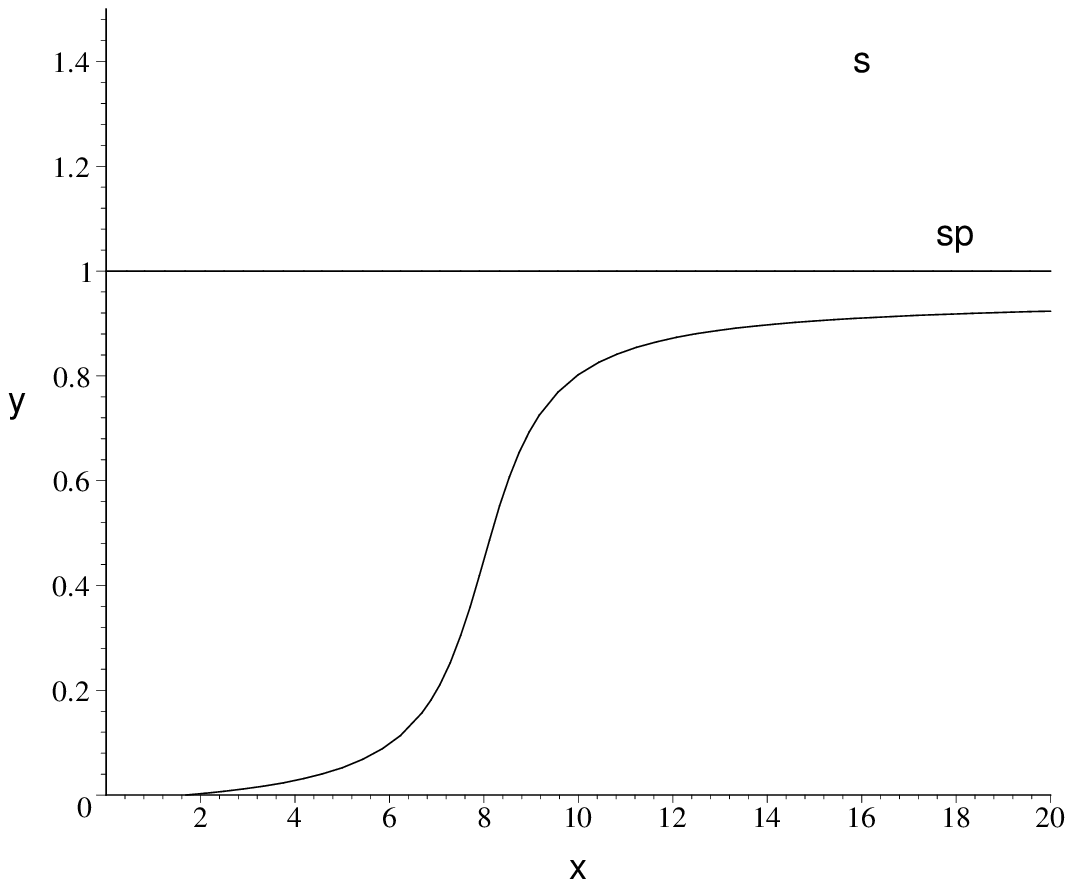}
  \caption{$\sigma(R)$: Case IIb when $\sigma''$ has one zero.}\label{fig:IIbBeta}
\end{figure}
\end{center}

\end{document}